\documentclass{raa}
\usepackage{graphicx,times}
\usepackage{upgreek}
\usepackage{natbib}
\bibpunct{(}{)}{;}{a}{}{,}

 \renewcommand\baselinestretch{1.22}
 \renewcommand{\thepage}{23--\arabic{page}}
\begin{document}
\title{Detecting exoplanets with FAST?}

\volnopage{{\bf 2019} Vol.~{\bf 19} No.~{\bf 2},~23(6pp)~
   {\small  doi: 10.1088/1674--4527/19/2/23}}
   \setcounter{page}{1}

\author{Philippe Zarka\inst{1}
\and Di Li\inst{2,3,4} \and Jean-Mathias Grie{\ss}meier\inst{5}
\and Laurent Lamy\inst{1} \and Julien N. Girard\inst{6} \and
\\[1mm] S\'{e}bastien L. G. Hess\inst{7} \and  T. Joseph W. Lazio\inst{8}
\and Gregg Hallinan\inst{9} }

\institute{LESIA, CNRS -- Observatoire de Paris, PSL 92190,
Meudon, France;
 {\it philippe.zarka@obspm.fr}\\
\and
National Astronomical Observatories, Chinese Academy of Science, Beijing 100101, China;
{\it dili@nao.cas.cn}\\
\and CAS Key Laboratory of FAST, National Astronomical
Observatories,
 Chinese Academy of Sciences, Beijing 100101, China\\
\and School of Astronomy and Space Science, University of Chinese
Academy of
Sciences, Beijing 101408, China\\
\and
LPC2E, CNRS -- Universit\'e d'Orl\'eans, 45071, Orl{\'e}ans, France\\
\and
Universit\'{e} Paris VII, AIM/Irfu/SAp/CEA-Saclay, Orme des
Merisiers, 91190, Gif-Sur-Yvette, France\\
\and
ONERA -- DESP, 31000, Toulouse, France\\
\and
 JPL -- California Institute of Technology, Pasadena, CA 91109, USA\\
\and
California Institute of Technology, Pasadena, CA 91125, USA\\
\vs\no {\small Received 2018 February 6; accepted 2018 June 5}}

\abstract{
 We briefly review the various proposed scenarios that may
lead to nonthermal radio emissions from exoplanetary systems
(planetary magnetospheres, magnetosphere-ionosphere and
magnetosphere-satellite coupling, and star-planet interactions),
and the physical information that can be drawn from their
detection. The latter scenario is especially favorable to the
production of radio emission above 70\,MHz. We summarize the
results of past and recent radio searches, and then discuss FAST
characteristics and observation strategy, including synergies. We
emphasize the importance of polarization measurements and a high
duty-cycle for the very weak targets that radio-exoplanets prove
to be. }

\titlerunning{{\it P. Zarka et al}.: ~Detecting Exoplanets with
FAST?}
\authorrunning{{\it P. Zarka et al}.: ~Detecting Exoplanets with FAST?}
\maketitle

\section{Introduction}

Planets are the most favorable cradle of %{for}
 life. As of today,
nearly 4000 exoplanets are known ({\it exoplanet.eu} -- mainly by
radial velocity or transit measurements, from which masses,
orbital parameters, sizes, densit{ies} and atmospheric
composition{s} can be inferred). In our solar system, magnetized
planets are strong radio sources (Jupiter is as bright as the Sun
at decameter wavelengths). Radio detection of exoplanets aims at
the physical characterization of exoplanets and comparative
studies with solar system planets.

\section{Experience from solar system planets and theory}

\begin{figure*}  %%fig1
\centering
\includegraphics[width=12cm]{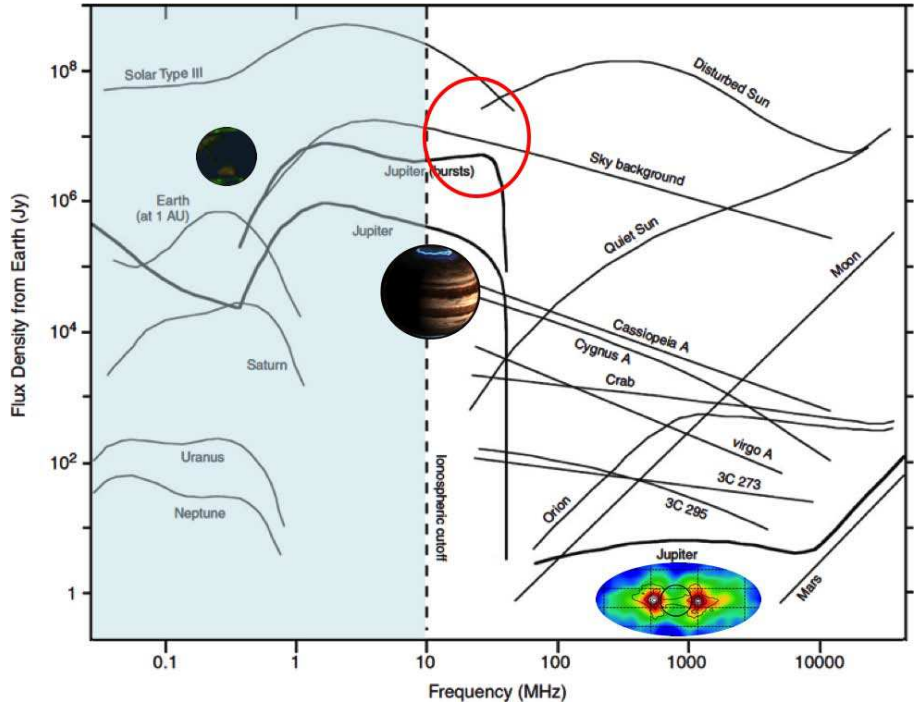}

\caption{\baselineskip 3.8mm Spectra of astronomical radio{
}sources detected {in} the Earth's vicinity. Auroral
planetary spectra lie below the Earth's ionospheric cut-off, except
Jupiter's decametric-to-decimetric emissions. Normalized to the same
observer distance of 1\,AU, Jupiter's spectrum must be upscaled by
$\times$20, Saturn's by $\times$100, Uranus' by $\times$400
and Neptune's by $\times$900, so that all are grouped within 2--3
orders of magnitude.\label{Fig1}}
\end{figure*}

There are {six} magnetized planets in the solar system
with{ a} planetary-scale magnetic field: Mercury, Earth, Jupiter, Saturn, Uranus and Neptune. In their
magnetospheres, various processes accelerate electrons to keV-MeV
energies, leading to high-latitude (auroral) radio emissions
(\citealt{Zarka+1998}), whose spectra are displayed in
Figure~\ref{Fig1}. The corresponding radio sources have been studied
remotely and in situ (e.g. \citealt{Huff+etal+1988};
\citealt{Treumann+Pottelette+2002}; \citealt{Lamy+etal+2010}). The
emissions were found to be coherent cyclotron ({m}aser)
radiation from keV electrons. The emission frequency depends on the
local cyclotron frequency, {which is }proportional to the
magnetic field amplitude, and is thus generally below a few
10s of MHz, reaching 40\,MHz for Jupiter. Emissions are very
intense (brightness temperature up to 10$^{15-20}$\,K), sporadic
(bursts lasting from msec to hours), anisotropic (beamed at large
angle{s} from the magnetic field) and circularly
polarized ({r}ight-{h}anded in northern
magnetic hemispheres and {l}eft-{h}anded in
southern ones) (\citealt{Wu+1985}; \citealt{Zarka+1998};
\citealt{Treumann+2006}; \citealt{Hess+etal+2008}). Weaker
incoherent synchrotron emission is generated by MeV electrons in
Jupiter's radiation belts (\citealt{dePater+1990}), {which was
}recently imaged with LOFAR (\citealt{Girard+etal+2016}).

The energy drivers for electron acceleration up to keV energies
include (\citealt{Zarka+etal+2001}; \citealt{Zarka+2007,
Zarka+2017}; \citealt{Nichols+2011}):

\begin{itemize}
\item[--]{Stellar {w}ind-{m}agnetosphere interaction (super-Alfv{\'{e}}nic, via compressions and reconnections),}
\item[--]{Magnetosphere-{i}onosphere coupling,}
\item[--]{Magnetosphere-{s}atellite coupling (sub-Alfv\'{e}nic, via reconnection or unipolar inductor interaction),}
\item[--]{Star-{p}lanet {i}nteraction (SPI -- sub-Alfv{\'{e}}nic, via reconnection or unipolar inductor interaction).}
\end{itemize}

\section{Predictions for exoplanets}

Jupiter's decametric emission is detectable at no more than
0.2\,pc over the Galactic background. Thus, the search for
exoplanetary radio emissions %{had to}
must had relied  %rely
 on scaling laws and extrapolations, as well
as more direct theoretical predictions.

\begin{figure*}  %%fig2
\centering
\includegraphics[width=16cm]{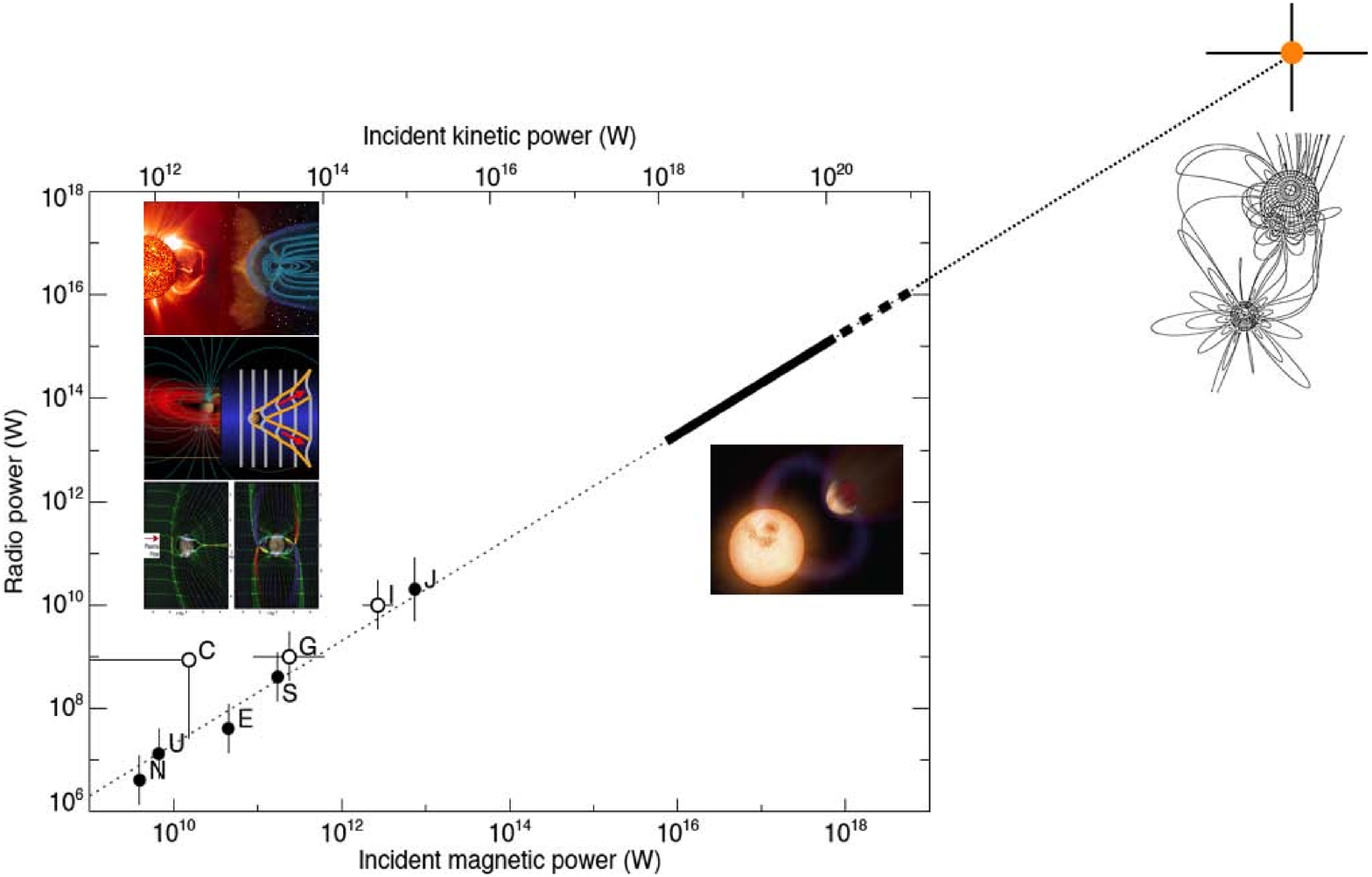}

\caption{\baselineskip 3.8mm Scaling law relating magnetospheric
(Earth, Jupiter, Saturn, Uranus and Neptune) and satellite-induced
(Io, Ganymede{ and} Callisto) radio power to incident Poynting
flux of the plasma flow on the obstacle. The {\it dashed line} has
a slope of 1, emphasizing
%the proportionality between ordinate and abscissa,
%
the proportionality between incident power and outgoing radio
power with a coefficient of 2$\times$10$^3$. The {\it thick bar}
extrapolates the magnetospheric interaction ({\it solid}) and
satellite-planet electrodynamic interactions ({\it dashed}){ to
hot Jupiters}. The {\it orange dot} {indicates} the case of the
RS\,{\,}CVn magnetic binary V711 $\tau$ discussed in
\citet{Zarka+2010}; \citet{Zarka+etal+2015a}. The inset{s}
{illustrate} the types of interaction.\label{Fig2}}
\end{figure*}

Study of solar wind-magnetosphere interactions led
\cite{Zarka+etal+2001} and \cite{Zarka+2007, Zarka+2010} {to
}propose that a planet's low-frequency radio output is
proportional to kinetic (including CME -- coronal mass ejection)
and magnetic (Poynting flux) power inputs on the obstacle's
cross-section. Extrapolation to hot {J}upiters led to prediction
of radio fluxes up to 10$^{3-5}$ times that of Jupiter
(Fig.~\ref{Fig2}). The Io-Jupiter electrodynamic interaction
transposed to plasma SPI led the same authors to predict radio
outputs proportional to the Poynting flux input, up to 10$^6$
times Jupiter's (Fig.~\ref{Fig2}).

\cite{Nichols+2011, Nichols+2012}, based on the
physics of magnetosphere-ionosphere interaction, predicted a large
low-frequency radio output, up to 10$^4$ times Jupiter's, for fast
rotating planets orbiting stars with a bright
{ultraviolet (}UV{)-X-ray} luminosity.
\cite{Willes+Wu+2004, Willes+Wu+2005} extended the theoretical frame
of{ the} Io-Jupiter unipolar interaction to terrestrial planets
around {w}hite {d}warfs, predicting a large
radio output at frequencies $\gg 1${\,}GHz.

In all cases, detection is the difficult step. Even if star-planet
systems will be unresolved in radio, subsequent discrimination
between stellar and planetary emission will rely (easily) on the
polarization{s} (circular for planets) and periodicities
(rotation, orbital) of the detected radio emission.

\begin{figure*}  %%fig3

\centering
\includegraphics[width=8.5cm]{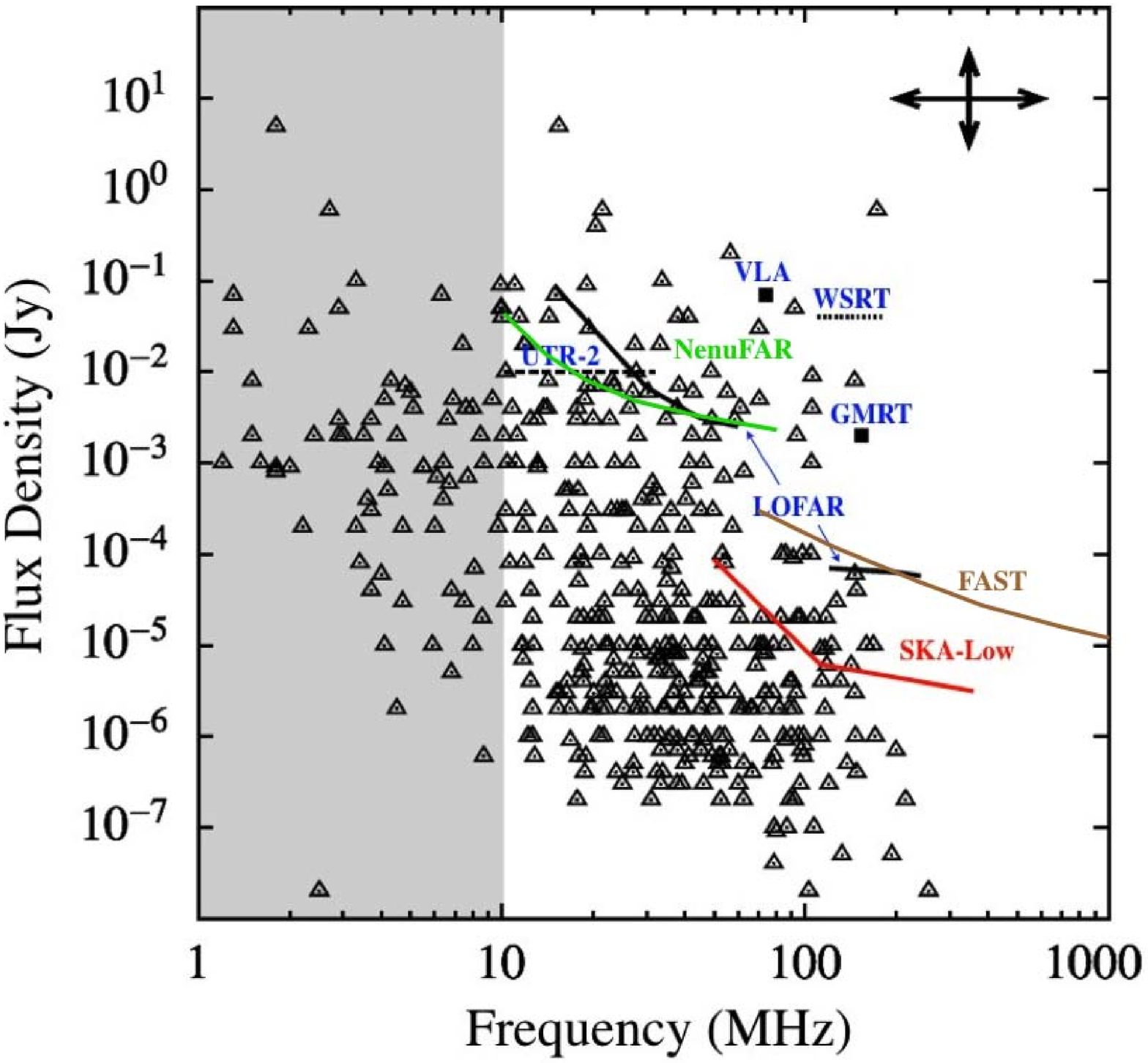}

\caption{\baselineskip 3.8mm Predicted maximum emission frequency
and expected radio flux for known exoplanets (in 2011, indicated
by {\it triangle symbols}) for a rotation-independent planetary
magnetic field and the scaling law {displayed in}
Figure~\ref{Fig2}. The approximate sensitivities of several
instruments {are} shown (for 1\,h of integration time and a
bandwidth of 4\,MHz). Frequencies below 10\,MHz are not observable
from the ground (ionospheric cut-off){;} {a}dapted from
(\citealt{Griessmeier+etal+2011}). More recent observations have
populated this diagram, but only $\sim$doubling the number of FAST
candidates.\label{Fig3}}\vs
\end{figure*}

\section{Motivations for studying radio emissions{ from exoplanets}}

Magnetospheric radio emissions provide unique information on
planetary magnetic field amplitude and tilt (Jupiter's magnetic
field was first measured that way
(\citealt{Burke+Franklin+1955})), and thus on the planetary
dynamo, giving constraints on the planetary interior structure. It
is also a signature of planetary rotation, due to the emission
anisotropy (the rotation periods of Jupiter
(\citealt{Higgins+etal+1997}), Saturn
(\citealt{Desch+Kaiser+1981}), Uranus and Neptune were measured
via their radio emission), that will permit directly testing
exoplanetary spin-orbit locking. It may reveal the presence of
satellites (e.g. \citealt{Bigg+1964}) interacting with the
planetary magnetic field. \cite{Hess+Zarka+2011} showed that radio
emissions also allow {researchers }to probe planetary orbit
inclination, and SPI and magnetospheric dynamics at large.

A magnetosphere also shields a planet's atmosphere and surface
{against} cosmic rays, stellar wind and CME
bombardment, preventing O$_3$ destruction and atmospheric erosion or
escape. It is thus a favorable condition for life
(\citealt{Griessmeier+etal+2004}; \citealt{Lammer+etal+2006}).
Finally{,} radio emission detection may evolve as an independent
discovery tool for planets around active, magnetic or variable
stars.

\section{Past and ongoing observations and results}

Targeted searches have been guided by theory and scaling laws
applied to the exoplanet census (with e.g. $\tau$\,{\,}Boo,
$\upsilon$~And or 55 Cnc as good candidates
(\citealt{Lazio+etal+2004}; \citealt{Griessmeier+etal+2007}),
strongly magnetized stars (such as HD\,{\,}189733
(\citealt{Donati+etal+2006})), planets with very elliptical orbits
and close-in periastron{s} (such as HD\,{\,}80606
(\citealt{Lazio+Farrell+2007}; \citealt{Lazio+etal+2010})) and
systems with possible optical SPI signatures (such as
HD\,{\,}179949 (\citealt{Shkolnik+etal+2008}) and references
therein). Observations were conducted with the VLA at
$\geq$74\,MHz (\citealt{Farrell+etal+2003, Farrell+etal+2004};
\citealt{Lazio+Farrell+2007}), UTR-2/Kharkov in the range
10--32\,MHz (\citealt{Ryabov+etal+2004}) and the GMRT at
$\geq$150\,MHz (\citealt{Hallinan+etal+2013};
\citealt{LecavelierdesEtangs+etal+2013}). No confirmed detection
{has been} reached. A hint of a radio occultation of Hat-P-11 b by
its star was found %at
 {by} GMRT, but in only one of two
observations with similar geometry, thus it remains to be
confirmed (\citealt{LecavelierdesEtangs+etal+2013}). Ongoing
targeted observations include campaigns at UTR-2 (10--32\,MHz,
$\geq$100\,h), %hours),
 programs with LOFAR in cycles 0 to 8
(20--80\,MHz, for a total of $\sim$170\,h), and a program with the
Long Wavelength Array (LWA) that plans to observe for
$\sim5000$\,h.

In parallel, correlations of the exoplanet catalog at exoplanet.eu
with low-frequency surveys open new perspectives. Four candidates
were found in the GMRT TGSS 150\,MHz survey, at the 10--100\,mJy
level, out of 175 exoplanetary systems in the surveyed field
(\citealt{Sirothia+etal+2014}), but these cases remain
unconfirmed. Analysis of LOFAR surveys is ongoing, with
Multi-Frequency Snapshot Sky Survey in the %120--
160\,MHz
band, at $\sim$5\,mJy sensitivity, and later in the 30--75\,MHz
band at $\sim$15\,mJy sensitivity from the LOFAR deep surveys
program (\citealt{Shimwell+etal+2017}; Loh et al. in preparation).
Permanent all-sky observations at $\sim$1\,s resolution have
started at{ the} Owens Valley Long Wavelength Array (OLWA).

As a preliminary conclusion, radio emissions much stronger than
Jupiter's at frequencies $\geq$150\,MHz appear to be rare, which
could be due to a too low planetary magnetic field, emission beaming
out of the observer's line of sight at the time of the
observations or too weak flux density.

\section{Science outcome enabled by FAST}

Following the previous conclusion, it is necessary to explore a
large sample of targets with the highest possible continuum
sensitivity, preferably at low frequencies, with circular
polarization or full Stokes observations, down to the thermal
noise level. Multi-epoch observations with integration times of a
few hours at each epoch are required for reaching a good
sensitivity and addressing time variations of intrinsically (or
beaming-induced) sporadic emissions.

\begin{table*}%[h!!]  %%table1
\centering
\begin{minipage}{7.5cm}
\caption{FAST Characteristics at Frequencies below L-band\label{Tab1}}
\end{minipage}

\fns\renewcommand\baselinestretch{1.35}
\begin{tabular}{lcccccc}%{lllllll}
\hline\noalign{\smallskip}
Frequency & $T_{\rm sky}$ &  $T_{\rm sys}$ & SEFD & Sensitivity
[4{\,}MHz$\times$1{\,}h, polarized] & Confusion$^{a}$[unpolarized] &  Angular resolution\\
(MHz)&  (K)& (K) & (Jy) & (mJy) & (mJy) & %(')
(arcmin)\\
\hline\noalign{\smallskip}
70&2500&100&110&0.5&3100&50\\
140&420&80&22&0.1&480&25\\
280&72&40&5&0.020&73&13\\
560&13&10&1&0.004&11&7\\
1150&5&10&0.7&0.003&1.6&3\\
\hline\noalign{\smallskip}
\end{tabular}
\parbox{140mm}{Notes: $^a$ Theoretical estimate from
\cite{Condon+2005}.}
\end{table*}

The FAST radio{ }telescope (\citealt{Nan+etal+2011}) can observe
down to 70\,MHz, within a broad frequency range, with full
polarization measurements. Table~\ref{Tab1} summarizes its
relevant characteristics for SPI studies. Comparing the
sensitivities with the typical flux densities of Jupiter's bursts
at 30--40{\,}MHz, about 40\,$\upmu$Jy at a 10\,pc range, it
appears that FAST has the sensitivity to detect moderate intensity
exoplanetary emissions, provided that the frequency of emission
exceeds 70\,MHz. A frequency $>$70\,MHz corresponds to cyclotron
emission from a source with a magnetic field amplitude $>25$\,G.
Figure~\ref{Fig3} shows the FAST sensitivity (as well as that of
other low-frequency instruments) compared to the predicted maximum
emission frequency and expected radio flux for known exoplanets in
2011, extrapolated following \cite{Griessmeier+etal+2007}. Only a
few candidate exoplanets have a predicted magnetic field reaching
or exceeding 25\,G. FAST will thus likely be best adapted to
search for SPI emissions (exoplanet-induced as proposed by
\cite{Zarka+2007, Zarka+2017}, including for terrestrial planets
around white dwarfs as proposed by \cite{Willes+Wu+2004,
Willes+Wu+2005}), down to moderate intensities, whereas
magnetospheric exoplanetary emissions will be rather the target of
lower frequency radio arrays such as LOFAR
(\citealt{vanHaarlem+etal+2013}) and NenuFAR
(\citealt{Zarka+etal+2012, Zarka+etal+2015b}).

Due to the relatively low angular resolution of FAST at low
frequencies, exoplanet detection via direct imaging will be
severely limited by confusion, i.e. spatial noise across the sky
background. The possible ways to lower this limit are (i)
observing in circular polarization --- for which confusion noise
is expected to be much smaller than for unpolarized emission ---
or full Stokes, and (ii) looking for time variable emission in
well-defined fields. The latter method is favored by FAST{'s}
constant beam size in all pointing directions
(\citealt{Nan+etal+2011}) and a frequency range well above the
ionospheric cut-off ($\sim$10\,MHz).

A key element of the success of FAST in exoplanet radio search and
study will be the possibility to conduct a large number of
multi-epoch targeted observations as well as extensive polarized
surveys (e.g. of all observable stellar systems up to{ a
distance of} 10\,pc: $\sim$200 known stars and
$\sim$35 currently known exoplanets). One drawback is the
availability of a single beam at low frequencies,
{which} will require a strong programmatic decision in
favor of exoplanet studies in order to perform a significant
program (such as the LWA -- HJUDE program). Follow-{up}
observations of targets identified by other instruments (LOFAR,
GMRT) also offer good prospects (FAST covers a large declination
range in common with these {two}
{facilities}).

\section{Synergies}
Advantage will be taken of synergies between radio observations of
exoplanets and of stellar solar-like bursts. Stellar flares could be planet induced or intrinsic: GHz
periodic bright pulses with periods $\sim$2\,h and 100\%
circularly polarized have been found by \cite{Hallinan+etal+2007,
Hallinan+etal+2008, Hallinan+etal+2015} and attributed to
cyclotron masers in{ a} brown dwarf with magnetic field
$\sim$2\,kG. Further, there is a {possibility} to
explore{ targets ranging} from brown dwarfs to exoplanets
(faster rotation, cooler {and} more neutral atmosphere,
{with }larger-scale stable magnetic field topologies{ and} weaker field amplitude). Tracing magnetic fields and
radio flux densities from brown dwarfs to planets will
{introduce} unique constraints for dynamo theories
and radio emission scaling laws. Lower mass planets seem to be more frequent
around M dwarfs, with close-in planets lying in the habitable zone
(\citealt{Dressing+Charbonneau+2013}). Commensal SETI searches would
be possible with adequate instrumentation (very high spectral
resolution of waveform measurements).

Synergies will also exist with observations at other wavelengths:
\begin{itemize}
\item[--]{with radio observations at frequencies below 70\,MHz (LOFAR, NenuFAR, UTR-2),}
\item[--]{with Zeeman Doppler Imaging (CFHT/Espadons, TBL/Narval, CFHT/Spirou) permitting stellar magnetic field measurements and planet searches around M dwarfs (\citealt{Fares+etal+2010}),}
\item[--]{with UV-X{-ray} observations ({\it HST}, {\it JWST}, {\it XMM-Newton}, {\it Chandra}, {\it Athena}{)} of stellar flares and exoplanet atmospheres,}
\item[--]{with {{\it PLATO}} and {{\it TESS}} revealing more nearby exoplanets,{ and} ESO-VLT/NGTS, ESPRESSO or {{\it Gaia}} surveys that will provide tens of exoplanets per {field-of-view} or a few degrees.}
\end{itemize}

\section{Conclusions}

Study of radio emissions from exoplanets and the star-planet
connection is a broad new field to explore. Its theoretical frame
is ready. There are currently optimistic prospects with LOFAR,
%LOFAR,
and its giant
 %and soon (in 2018) with its giant
 compact
extension NenuFAR. The advent of SKA will {enable} a sensitivity
improvement of $\sim$30$\times$ compared to LOFAR, down to the
10\,$\upmu$Jy level, at frequencies $\geq$50\,MHz. This make{s} it
highly likely that SKA-Low will detect exoplanetary radio signals
(\citealt{Zarka+etal+2015a}). But this will happen in 2020+, and
FAST, being operational now, benefits from a favorable window for
exoplanet and SPI radio search. It will also cover a declination
range complementary to that of LOFAR {and} NenuFAR (mostly the
northern hemisphere),{ and} SKA (mostly the southern hemisphere)
and{ be} compatible with that of the GMRT.

%\vs

\begin{acknowledgements}
This work is supported by the National Key R\&D Program No.
2017YFA0402600, the CAS International Partnership Program No.
14A11KYSB20160008, and the NSFC grant No. 11725313.
Part of this research (TJWL) was carried out at
JPL/Cal{t}ech under a contract with NASA.
\end{acknowledgements}

\end{document}